  \documentclass[prl,aps,twocolumn,showpacs]{revtex4}
\usepackage[dvips]{graphicx}
\usepackage{amsmath,  amssymb, array, graphics,  amsfonts}
\usepackage{dcolumn}
\begin{document}
\topmargin=-15mm

\title {\bf
 Skin effect with arbitrary specularity in Maxwellian plasma}

\author{\bf Anatoly V. Latyshev and Alexander Yushkanov}

\affiliation  {Department of Mathematical Analysis and
Department of Theoretical Physics, Moscow State Regional
University,  105005, Moscow, Radio st., 10--A}

\begin{abstract}
The problem of skin effect with arbitrary specularity in
maxwellian plasma with specular--diffuse boundary conditions is
solved. New analytical method is developed that makes it
possible to obtain a solution up to an arbitrary degree of
accuracy. The method is based on the idea of symmetric
continuation of not only the electric field, but also electron
distribution function. The solution is obtained in a form of von
Neumann series.

\noindent {\bf Keywords: skin effect, specular--diffuse boundary
conditions, analytical method, von Neumann series.}
   \end{abstract}

\pacs{52.35.-g, 52.2.-j, 52.25.-b}
\date{\today}
\maketitle

\section{I. Introduction}

The skin effect problem is one of the most important problems in
plasma kinetic theory  (see, for example, in Refs.\cite{1}--\cite{6}).
The skin effect in plasma is a response of electron gas to
external transverse electromagnetic field.
The problem also has great practical importance.

The solution of the skin effect problem with specular reflection
boundary conditions is well--known \cite{1}, \cite{2}. The
analytic solution of the problem with diffuse reflection
boundary conditions has been obtained in the middle of the previous
century (see, for example, Ref.\cite{4}).

The skin effect problem with general specular -- diffuse
reflection boundary conditions \cite{6} is not solved till now.
It's well known that specularity coefficient $q$ is a very
important factor in the kinetic skin effect theory \cite{7} --
\cite{9}. The limiting cases $q=0$ (diffuse surface scattering
of electrons) and $q=1$ (specular surface scattering of
electrons) are only very special cases. Actually the specularity
coefficient $q$ equals neither to zero, nor unit, and
takes some intermediate values on.

So, for example, in the work \cite{6} it is shown, that the
specularity coefficient $q$ is equal to $0.4$ in $Na$ wire. In this
connection  the skin effect problem   with specular--diffuse
boundary conditions has exclusively fundamental significance. Its value is
great for theories, and for practical applications. So it's
obvious that the solution of the skin effect problem with
general specular--diffuse reflection boundary conditions is a
very important task.

The method of solution of this problem for degenerate plasma in
metal has been developed in Ref. \cite{8}. This method is based on the use
of von Neumann series. Authors in Ref. \cite{8} have demonstrated
high efficiency of the method developed for computation of
the skin effect characteristics. The goal of this work is
generalization of the method developed in Ref. \cite{8} in the case
of gaseous plasma.

By method of decomposition of the solution by eigenfunctions of
the corresponding characteristic equation exact solutions of the
skin effect problem in metal for diffuse and specular boundary
conditions
 are received in Refs. \cite {9}, \cite{10}.

In the last years interest to skin effect problems continues to grow
(see, for example, Refs. \cite{11}--\cite{16}). In particular, in
Ref. \cite{16} in limiting anomalous skin effect conditions, the
oblique electromagnetic wave reflection from the sharp plasma
boundary in an assumption of mixed (specular and
diffuse) electron reflection from the boundary is considered.

\section{II. Problem statement }

Let's gaseous (nondegenerate) plasma occupy a half--space
$x>0$. The distribution function of electrons
$f=f(t,\mathbf{r},\mathbf{v})$ is normalized by electron  numerical
density (concentration of electrons):
$$
\int f(t,\mathbf{r},\mathbf{v})\,d^3v=n(t,\mathbf{r}),
$$
where $\mathbf{p}=m \mathbf{v}$ is the electron momentum, $m$
is the electron mass, $e$ is the electron charge, $d^3v=dv_xdv_ydv_z$.

We consider electromagnetic wave which propagates in
direction orthogonal to the plasma surface. Then the external
field has only one $y$--component. The internal field inside
plasma has only $y$--component
$$
E_y(t,x)=e^{-i \omega t}E(x)
$$
too, where $\omega$ is the field frequency.

To describe the electron distribution function we will use the
Vlasov --- Boltzmann kinetic equation. The collision integral
will be represented in the form of $\tau$--model
$$
\frac{\partial f}{\partial t}+v_x\frac{\partial f}{\partial
x}+eE_y(x,t)\frac{\partial f}{\partial p_y}=
\frac{f_0(v)-f(x,v_x,t)}{\tau},
$$
where $\tau$ is the time between two electrons collisions,
$\tau=1/\nu$, $\nu$ is the effective electron collision
frequency, $f_0(v)$ is the equilibrium maxwellian distribution
function,
$$
f_0(v)=n\Big(\frac{m}{2\pi k_B
T}\Big)^{3/2}\exp\Big(-\frac{mv^2} {2k_BT}\Big).
$$

For weak fields this equation may be linearized:
$$
f=f_0(C)\Big(1+ C_y h(x, C_x)e^{-i\omega t}\Big).
\eqno{(1.1)}
$$

Here
$$
f_0(C)= n\Big(\frac{\beta}{\pi}\Big)^{3/2}\exp \big(-C^2\big),
$$
where $\beta=m/(2k_BT)$, $\mathbf{C}=\sqrt{\beta}\mathbf{v}$,
$\mathbf{C}$ is the dimensionless electron velocity,
$k_B$ is the Boltzmann costant, $T$ is the plasma temperature.

For function $h(x, C_x)$ we have the following kinetic equation:
$$
\mu\frac{\partial h(x_1,\mu)}{\partial
x_1}+z_0\,h(x_1,\mu)=e(x_1),
\eqno{(1.2)}
$$
where
$$
 z_0=1-i\frac{\omega}{\nu}.
$$

In the equation (1.2) $x_1$ is the dimensionless coordinate,
$$
\mu=v_x\sqrt{\beta},   \quad x_1=\frac{x}{l}=\nu \sqrt{\beta}x,\quad
t_1=\nu\,t,
$$
$t_1$ is the dimensionless time, $l$ is mean
electron free path and $e(x_1)$ is the dimensionless electric
field:
$$
e(x_1)=\frac{\sqrt{2}\,e}{\nu\sqrt{mk_BT}}\,E(x_1).
$$

We will neglect displacement current. Then the equation for
electric field may be written in the form:
$$
{e''}(x_1)
=-\frac{4\pi i \omega}{c^2}\,j_y(x_1), \eqno{(1.3)}
$$
where $j_y(x_1)$ is the electric current,
$$
{j}_y(x_1)=e\int {v}_y [f_0(v)+\sqrt{\beta}v_ye^{-i\omega t}
h(x,v_x)]\,d^3v.
\eqno{(1.4)}
$$

Let's  extend the electric field and the electric distribution
function on the "negative"\, half--space $x<0$ in  symmetric
manner. For the functions $E(x)$ and $h(x,\mu)$ we will then write:
$$
{e}(x_1)={e}(-x_1), \;\qquad h(x_1,\mu)=h(-x_1,-\mu).
\eqno{(1.5)}
$$

We may rewrite the equation (1.4) with use of dimensional parameter
$\alpha$:
$$
\frac{d^2e(x_1)}{dx_1^2}
=-i\frac{\alpha}{\sqrt{\pi}}
 \int\limits_{-\infty}^{\infty} \exp(-{\mu'}^2)h(x_1,\mu')\,d\mu'.
\eqno{(1.6)}
$$

Here
$$
\alpha=\frac{4 \pi e^2n\omega} {c^2\,\beta\nu^3m}=2
\Big(\frac{l}{\delta}\Big)^2, \quad \delta^2=\frac{c^2}{2\pi
\omega \sigma},
$$
$$
\sigma=\frac{e\,n}{m\,\nu},\quad
l=v_T\tau, \quad v_T=\frac{1}{\sqrt{\beta}},
$$
$v_T$  is the electron thermal velocity, $\delta $ is the
penetration length of external electric field for normal
skin effect, $\sigma $ is the plasma conductivity.

By extension procedure (1.5) on the half--space $x<0$ we may
include the surface conditions in the equation for skin effect
problem.

Specular -- diffuse boundary conditions on the boundaries of
positive and negative half--spaces may be written in the form:

$$
h(+0,\mu)=qh(+0,-\mu), \quad 0<\mu<1,
$$
$$
h(-0,\mu)=qh(-0,-\mu), \quad -1<\mu<0,
$$
where $q$ is the specularity coefficient, $0 \leqslant q \leqslant 1$.

In accordance with (1.5) we obtain:
$$
h(+0,\mu)=q h(-0,\mu), \qquad 0<\mu<1,
\eqno{(1.7)}
$$
$$
h(-0,\mu)=q h(+0,\mu), \qquad -1<\mu<0.
\eqno{(1.8)}
$$

The required function $h(x_1,\mu)$ and the electric field must decay
away from the surface:

$$
h(+\infty,\mu)=0, \qquad e(+\infty)=0. \eqno{(1.9)}
$$

We assume that the gradient of the electric field is finite and
known at the plasma boundary:

$$
e'(0)=e_s',\qquad |e_s'|<+\infty.
\eqno{(1.10)}
$$

Here, the gradient of the electric field on the plasma boundary
$e_s'$ is given.

\section{III. Characteristic system}

The variable $x_1$ will be denoted again  by $x$.

Let's include boundary conditions (1.7) and (1.8) in the kinetic
equation (1.2), and boundary condition (1.10) include in the
electric field equation (1.6).

As a result we will obtain system of equation  for skin effect in
half--space of the plasma:
$$
\mu\frac{\partial h}{\partial x}+z_0\,h(x,\mu)=
$$
$$
e(x)- (1-q)|\mu|
\,h(\mp 0,\mu)\delta(x),\quad \pm \mu>0, \eqno{(2.1)}
$$
$$
\frac{d^2 e(x)}{dx^2}
=-i\frac{\alpha}{\sqrt{\pi}}\int\limits_{-\infty}^{\infty}
\exp(-{\mu'}^2)h(x,\mu')\,d\mu'+
$$
$$
2e_s^{'}\delta(x).
\eqno{(2.2)}
$$

The impedance is determined by formula \cite{1}:
$$
Z=\frac{4\pi i \omega}{c^2}\frac{E_y(0)}{E_y'(0)},
$$

With the use of dimensionless field $e(x)$ this relation may be
rewritten in the form:
$$
Z=\frac{4\pi i \omega l}{c^2}\frac{e(0)}{e_s^{'}}.
$$

From equation (2.1)  and boundary conditions (1.9) we obtain the
following expression for $x>0,\; \mu<0$:
$$
h_{+}(x,\mu)=-\frac{1}{\mu}\exp \Big(-\frac{z_0x}{\mu}\Big)
\int\limits_{x}^{\infty} \exp\Big(-\frac{z_0
t}{\mu}\Big)\,e(t)\, dt.
$$

In the case $x<0,\; \mu>0$ we obtain:
$$
h_{-}(x,\mu)=\frac{1}{\mu}\exp\Big(-\frac{z_0x}{\mu}\Big)\,
\int\limits_{-\infty}^{x} \exp\Big(-\frac{z_0
t}{\mu}\Big)\,e(t)\, dt.
$$

Then we may rewrite the equation (2.1) in the form:
$$
\mu\frac{\partial h}{\partial x}+z_0h(x,\mu)-e(x)=
$$
$$
- (1-q)|\mu|
\,h_{\pm}(0,\mu)\delta(x),\quad \pm \mu>0.
\eqno{(2.3)}
$$

The solution of the equations (2.2) and (2.3) we will seek in the
form of Fourier integrals:
$$
e(x)=\frac{1}{2\pi}\int\limits_{-\infty}^{\infty}
e^{ikx}E(k)\,dk,
\eqno{(2.4)}
$$
$$
\delta(x)=\frac{1}{2\pi}\int\limits_{-\infty}^{\infty}
e^{ikx}\,dk,
$$
$$
h(x,\mu)=\frac{1}{2\pi}\int\limits_{-\infty}^{\infty}
e^{ikx}\Phi(k,\mu)\,dk. \eqno{(2.5)}
$$

Then for the function $h_{+}(x,\mu)$ the following  expression
may be derived:
$$
h_{+}(x,\mu)=-\frac{\exp(-z_0x/\mu)}{2\pi\mu} \times
$$
$$
\times
\int\limits_{-\infty}^{\infty}dk \int\limits_{x}^{\infty}dt
\exp\Big(ikt+\frac{z_0t}{\mu}\Big)E(k)=
$$
$$
=\frac{1}{2\pi}\int\limits_{-\infty}^{\infty}
\frac{\exp(ikx)E(k)}{z_0+ik\mu}dk.
\eqno{(2.6)}
$$

It's may be proved, that the expression for $h_{-}(x,\mu)$
coincides with the expression for $h_{+}(x,\mu)$. Therefore  we
have
$$
h_{\pm}(0,\mu)=\frac{1}{2\pi}
\int\limits_{-\infty}^{\infty}\frac{E(k)\,dk}{z_0+ik\mu}.
$$

We substitute the expressions (2.4), (2.5) and (2.6) into the equations
(2.2) and (2.3). This procedure leads to characteristic
system of equations:
$$
\Phi(k,\mu)(z_0+ik\mu)=
$$
$$
E(k)-(1-q)\frac{|\mu|}{2\pi}
\int\limits_{-\infty}^{\infty}\frac{E(k_1)\,dk_1}{z_0+ik_1\mu},
\eqno{(2.7)}
$$
$$
-k^2 E(k)=2e_s^{'}-
i\alpha\int\limits_{-\infty}^{\infty}\exp(-\mu^2)\Phi(k,\mu)\,d\mu.
\eqno{(2.8)}
$$

The function $e(x)$ is an even function. Then $E(-k)=E(k)$, and
equation (2.7) may be rewritten as
$$
\Phi(k,\mu)(z_0+ik\mu)=
$$
$$
E(k)-(1-q)\frac{|\mu|z_0}{\pi}
\int\limits_{0}^{\infty}\frac{E(k_1)\,dk_1}{z_0^2+k_1^2\mu^2}.
\eqno{(2.9)}
$$

Let's substitute the expression (2.9) into the equation (2.8). Then we
obtain:

$$
L(k)E(k)=-2e_s^{'}+$$
$$
(1-q)\frac{\alpha z_0^2}{\pi \, i}
\int\limits_{0}^{\infty}E(k_1)\,J(k,k_1)\,dk_1. \eqno{(2.10)}
$$

Here $L(k)$ is the dispersion function,

$$
L(k)=k^2-i\frac{\alpha}{\sqrt{\pi}}
\int\limits_{-\infty}^{\infty}\frac{\exp(-\mu^2)\,d\mu}{z_0+ik\mu}=
$$
$$
k^2-\frac{2i z_0\alpha}{\sqrt{\pi}}
\int\limits_{0}^{\infty}\frac{\exp(-\mu^2)\,d\mu}
{z_0^2+k^2\mu^2},
$$
and $J(k,k_1)$ is the integral
$$
J(k,k_1)=\frac{2}{\sqrt{\pi}}\int\limits_{0}^{\infty}
\frac{\exp(-u^2)\,du}{(z_0^2+k^2u)(z_0^2+k_1^2u)}.
$$

Characteristic system consists of two equations (2.9)
and (2.10).

The integral $J (k_1, k_2) $ we will express through the
integral exponential  function. We will spread out integrand
 partial fractions:
$$
\frac{1}{(z_0^2+k_1^2u)(z_0^2+k_2^2u)}=
\frac{A}{z_0^2+k_1^2u}+\frac{B}{z_0^2+k_2^2u},
$$
where
$$
A=-\frac{k_1^2}{z_0^2(k_2^2-k_1^2)}, \qquad
B=\frac{k_2^2}{z_0^2(k_2^2-k_1^2)}.
$$

Thus, we receive, that
$$
J(k_1,k_2)=A\int\limits_{0}^{\infty}\frac{e^{-u}\,du}
{z_0^2+k_1^2u}+B\int\limits_{0}^{\infty}\frac{e^{-u}\,du}
{z_0^2+k_2^2u},
$$
or
$$
J(k_1,k_2)=AJ_0(k_1)+BJ_0(k_2),
$$
where
$$
J_0(k_j)=\int\limits_{0}^{\infty}\frac{e^{-u}\,du}
{z_0^2+k_j^2u}=
$$
$$
\frac{1}{k_j^2}\exp(\frac{z_0^2}{k_j^2})
\int\limits_{z_0^2}^{\infty}\exp(-\frac{x}{k_j^2})\frac{dx}{x},
\quad j=1,2.
$$

\section{IV. Solution of the problem with the use of von Neumann series}

Let's expand the solution of equations (2.9), (2.10) by the
following series:
$$
E(k)=E_0(k)+(1-q)\,E_1(k)+(1-q)^2\,E_2(k)+\cdots,
\eqno{(3.1)}
$$
$$
\Phi(k,\mu)=
\Phi_0(k,\mu)+(1-q)\Phi_1(k,\mu)+
$$
$$
(1-q)^2\Phi_2(k,\mu)+\cdots.
\eqno{(3.2)}
$$

Functions $E_j(k)$ and $\Phi_j(k,\mu)\;(j=1,2,3,\cdots)$ may be
obtained from the characteristic system. For zero approximation we
have:

$$
E_0(k)=-\frac{2e_s^{'}}{L(k)},\quad
\Phi_0(k)=-\frac{2e'_s}{L(k)(z_0+ik\mu)}. \eqno{(3.3)}
$$

For first approximation we obtain:
$$
E_1(k)=\frac{ \alpha z_0^2}{L(k)\pi\, i}\int\limits_{0}^{\infty}
E_0(k_1)J(k,k_1)\,dk_1,
\eqno{(3.4)}
$$
$$
\Phi_1(k,\mu)=\frac{E_1(k)}{z_0+ik\mu}-\frac{z_0|\mu|}
{(z_0+ik\mu)\pi}\int\limits_{0}^{\infty}\frac{E_0(k_1)\,dk_1}
{z_0^2+k_1^2\mu^2}.
\eqno{(3.5)}
$$

For $n$-approximation the following expression may be derived:
$$
E_n(k)=
$$
$$
\frac{\alpha z_0^2}{L(k)\pi\,i}\int\limits_{0}^{\infty}
E_{n-1}(k_1)J(k,k_1)\,dk_1, \quad n=1,2,\cdots, \eqno{(3.6)}
$$
$$
\Phi_n(k,\mu)=\frac{E_n(k)}{z_0+ik\mu}-
$$
$$
\frac{z_0|\mu|}
{(z_0+ik\mu)\pi}\int\limits_{0}^{\infty}\frac{E_{n-1}(k_1)\,dk_1}
{z_0^2+k_1^2\mu^2},\quad n=1,2,\cdots.
\eqno{(3.7)}
$$

We may rewrite the expressions (3.1) in the form
$$
E_1(k)=-2e_s^{'}\frac{\alpha z_0^2}{\pi\, i}
\int\limits_{0}^{\infty}\frac{J(k,k_1)}{L(k)L(k_1)}\,dk_1,
$$
$$
E_2(k)=-2e_s^{'}\Big(\frac{\alpha z_0^2}{\pi\, i}\Big)^2 \times
$$
$$
\times
\int\limits_{0}^{\infty}\int\limits_{0}^{\infty} \frac
{J(k,k_1)J(k_1,k_2)}{L(k)L(k_1)L(k_2)}dk_1dk_2,\cdots.
$$

In general case when $n=1,2,3, \cdots,$ we have:
$$
E_n(k)=-2e_s^{'}\Big(\frac{\alpha z_0^2}{\pi\, i}\Big)^n \times
$$
$$ \times
\int\limits_{0}^{\infty}\cdots \int\limits_{0}^{\infty}
\frac{J(k,k_1)J(k_1,k_2)\cdots J(k_{n-1},k_n)}{L(k)L(k_1)\cdots
L(k_n)}dk_1\cdots dk_n.
\eqno{(3.8)}
$$

Therefore the series (3.1) constructed may be expressed in
the explicit form:
$$
E(k)=-\frac{2e_s^{'}}{L(k)}\Bigg[1+\sum\limits_{n=1}^{\infty}
(1-q)^n\Big(\frac{\alpha z_0^2}{\pi i}\Big)^n \times $$$$ \times
\int\limits_{0}^{\infty}\cdots \int\limits_{0}^{\infty}
\frac{J(k,k_1)J(k_1,k_2)\cdots J(k_{n-1},k_n)}{L(k_1)\cdots
L(k_n)}dk_1\cdots dk_n\Bigg].
$$

\section{V. Electric field, distribution function and surface impedance}

In accordance with (2.4) and (2.5) we will construct expressions
for electric field and distribution function. Using the expressions
(3.1) and (3.2) we obtain:
$$
e(x)=\frac{1}{\pi}\sum\limits_{n=0}^{\infty}(1-q)^n
\int\limits_{0}^{\infty}E_n(k)\cos kx\,dk, \eqno{(4.1)}
$$

$$
h(x,\mu)=\frac{1}{\pi}\sum\limits_{n=0}^{\infty}(1-q)^n
\int\limits_{-\infty}^{\infty}e^{ikx}\Phi_n(k,\mu)\,dk.
\eqno{(4.2)}
$$

We rewrite the expression (4.1) with the use of (3.8) in the
following form:

$$
e(x)=-\frac{2e_s^{'}}{\pi}\int\limits_{0}^{\infty}\frac{ \cos
kx\,dk}{L(k)}\Bigg[1+\sum\limits_{n=1}^{\infty}
(1-q)^n\Big(\frac{\alpha z_0^2}{\pi\,i}\Big)^n \times
$$
$$
\times \int\limits_{0}^{\infty}\cdots \int\limits_{0}^{\infty}
\frac{J(k,k_1)J(k_1,k_2)\cdots J(k_{n-1},k_n)}{L(k_1)\cdots
L(k_n)}dk_1\cdots dk_n\Bigg].
$$

Function $h(x,\mu)$ may be written in the form:
$$
h(x,\mu)=\frac{1}{2\pi}\int\limits_{-\infty}^{\infty}
\Bigg[E(k)-
$$
$$(1-q)\frac{z_0|\mu|}{\pi} \int\limits_{0}^{\infty}
\frac{E(k_1)\,dk_1}{z_0^2+k_1^2\mu^2}\Bigg]
\frac{e^{ikx}dk}{z_0+ik\mu},
\eqno{(4.3)}
$$

Expression (4.3) may be written also in the next form:
$$
h(x,\mu)=\frac{1}{\pi}\int\limits_{0}^{\infty} \frac{z_0\cos
kx+k\mu \sin kx}{z_0^2+k^2\mu^2}\Big[E(k)-
$$
$$
-(1-q)\frac{z_0|\mu|}{\pi^2} \int\limits_{0}^{\infty}
\frac{E(k_1)\,dk_1}{z_0^2+k_1^2\mu^2}\Big]dk.
$$

If we know function $h(x,\mu)$ we may write down the electron
distribution function $f$ according to the equation (1.1).

Let us consider now the calculation of impedance:
$$
Z=\frac{4\pi i \omega l}{c^2}\frac{e(0)}{e_s^{'}}= \frac{2i
\omega l} {c^2e_s^{'}}\int\limits_{-\infty}^{\infty}E(k)\,dk=
$$
$$
\frac{4i \omega l}{c^2 e_s^{'}}\int\limits_{0}^{\infty}E(k)\,dk.
$$

We decompose $Z$ in the following series:
$$
Z=Z_0+(1-q)Z_1+(1-q)^2Z_2+\cdots. \eqno{(4.4)}
$$
Here
$$
Z_n=\frac{4i \omega l}{c^2e_s^{'}}\int\limits_{0}^{\infty}
E_n(k)\,dk, \quad n=0,1,2,\cdots.
\eqno{(4.5)}
$$

Now we write down expressions for zero, first and second
approximations for impedance (see (4.4) and (4.5))
$$
Z_0=-\frac{4i \omega l}{c^2}\int\limits_{-\infty}^{\infty}
\frac{dk}{L(k)}=-\frac{8i \omega l}{c^2}\int\limits_{0}^{\infty}
\frac{dk}{L(k)},
\eqno{(4.6)}
$$
$$
Z_1=-\frac{8i\omega l}{c^2}\frac{\alpha z_0^2}{\pi \, i}
\int\limits_{0}^{\infty}
\frac{dk}{L(k)}\int\limits_{0}^{\infty}\frac{J(k,k_1)\,dk_1}
{L(k_1)},
$$
$$
Z_2=-\frac{8i \omega l}{c^2}\Big(\frac{\alpha z_0^2}
{\pi\,i}\Big)^2 \int\limits_{0}^{\infty}\times
$$
$$
\times
\frac{dk}{L(k)}\int\limits_{0}^{\infty}\frac{J(k,k_1)dk_1}{L(k_1)}
\int\limits_{0}^{\infty}\frac{J(k_1,k_2)dk_2}{L(k_2)}.
$$

Expression for general term of series (4.4) has the form:
$$
Z_n=-\frac{8 i \omega l}{c^2}\Big(\frac{\alpha z_0^2} {\pi\,
i}\Big)^n \times
$$
$$ \times
\int\limits_{0}^{\infty} \cdots
\int\limits_{0}^{\infty}\frac{J(k,k_1)J(k_1,k_2) \cdots
J(k_{n-1},k_n)}{L(k)L(k_1)\cdots L(k_n)}\times
$$
$$
\times
dk\,dk_1\cdots dk_n.
$$

\section{VI. Analysis and discussion}

In the previous sections we have considered the method, leading to
exact solution of the skin effect problem with arbitrary specularity
coefficient. In the case $q=1$ the method leads to the classical
solution (4.6) of the problem with  specular surface
conditions (see, for example \cite{5}, \cite{10}, \cite{17}). In
\cite{10} this classical solution is represented in the form:
$$
Z_{\rm ref}=-\frac{8i \omega l}{c^2}\int\limits_{0}^{\infty}
\frac{d\tau}{\lambda(iz_0\tau)}, \eqno{(5.1a)}
$$
where
$$
\lambda(iz_0\tau)=1-
\alpha\,\tau^3
\int\limits_{-\infty}^{\infty}\frac{\exp(-\mu^2)\,d\mu}
{\mu-iz_0\tau}.
\eqno{(5.1b)}
$$

The comparison of the expressions (5.1) and (4.6) gives
$Z_{ref}=Z_0$. Indeed, after the change of variables in the
integral we have:
$$
\int\limits_{0}^{\infty}\frac{d\tau}{\lambda(iz_0\tau)}=
\int\limits_{0}^{\infty}\frac{d\tau}{\tau^2\,\lambda
\big(iz_0/\tau\big)}=\int\limits_{0}^{\infty}\frac{dk} {L(k)}.
$$

Now let us consider second approximation:
$$
Z=Z_0+(1-q)Z_1+(1-q)^2Z_2.
$$

When  $q=1$ this solution is exact. Maximum deviation from exact
solution corresponds to the case when $q=0$. The exact solution
of the problem in the case $q=0$ is also well known \cite{3},
\cite{10}:
$$
Z_{dif}=-\frac{4\pi i \omega l}{c^2}\Bigg[\frac{2}{\pi}
\int\limits_{0}^{\infty}
\ln\Big[\lambda(iz_0\tau)\Big]\frac{d\tau}{\tau^2}\Bigg]^{-1}.
\eqno{(5.2)}
$$

It's convenient to rewrite the expression (5.2) with the use of
our notations:
$$
Z_{dif}=-\frac{4\pi i \omega l}{c^2}\Bigg[\frac{2}{\pi}
\int\limits_{0}^{\infty} \ln\Big[k^{-2}L(k)\Big]\,dk\Bigg]^{-1}.
$$

Now consider the ratio of real (and imaginary) parts of the
solutions constructed  in zero, first and second approximations
to the solution in zero approximation $\Re (Z_0)$ ($\Im (Z_0)$)
for the case $q=0$. The last solution coincide with the solution
of the problem with specular scattering boundary conditions.

We will build two plots (curves 1 and 2):
$$
Y_1=\frac{\Re(Z_0+Z_1)}{\Re(Z_0)}=1+\frac{\Re(Z_1)}{\Re(Z_0)}
$$
and
$$
Y_2=\frac{\Re(Z_0+Z_1+Z_2)}{\Re(Z_0)}=1+\frac{\Re(Z_1)}{\Re(Z_0)}+
\frac{\Re(Z_2)}{\Re(Z_0)}
$$
or
$$
Y_2= Y_1+\frac{\Re(Z_2)}{\Re(Z_0)},
$$
and also analogous plots for the ratios of the imaginary parts.

Here $Z_0$ is the solution for the case of specular suface
conditions. Values $Z_1$ and $Z_2$ correspond to corrections for
the first and the second approximations.

The curves 3 on the plots correspond the ratios of impedance for
diffuse scattering surface condition to impedance for specular
scattering surface condition $\Re(Z_{dif})/\Re(Z_{ref})$
($\Im(Z_{dif})/\Im(Z_{ref})$).

The derived method has maximum error in the case of extremely
anomalous skin effect, when parameter $\alpha\gg 1$. In this
case ratios  defined above are equal to 1.125. So it is obvious
from the plots, that in zero approximation the method error is
equal to $12.5\%$.

For the first approximation in this case we have
$Z_{dif}/Z=1.03$. So the first approximation error is equal to
$3\%$. For the second approximation in this case we have
$Z_{dif}/Z=1.01$. And the second approximation error is equal to
$1\%$.

The analysis of plots shows, that the considered impedance
ratios in first approximation coincide with the exact solution
when $\alpha<10^{-1}$. For the second approximation the
coincidence is observed when $\alpha<1$.

\section{VII. Conclusion}

The effective method of the solution of boundary problems of
the kinetic theory  is developed. This method is based on symmetric
continuation of the electric field and distribution function of
electrons.

The offered method  gives an
error less, than $1 \% $ in the second approximation already.
The method is accurate and allows to
construct exact solution in the form of  von Neumann series.

\begin{figure}[h]
\begin{center}
\includegraphics[width=8.5cm, height=4.5cm]{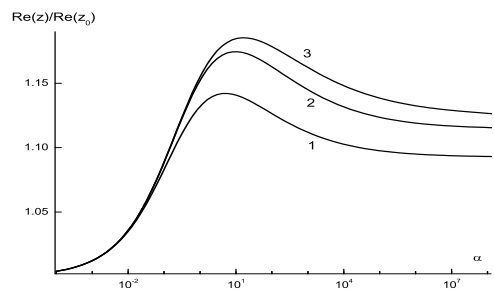}
\caption{Dependence of $\Re(Z)/\Re(Z_0)$ on parameter $\alpha$
for the case $q=0$. The curve $3$ corresponds to diffuse
scattering boundary conditions, the curves $1,2$ correspond to
first and second approximations.}\label{rateI}
\end{center}
\end{figure}

\begin{figure}[t]
\begin{center}
\includegraphics[width=8.5cm, height=4.5cm]{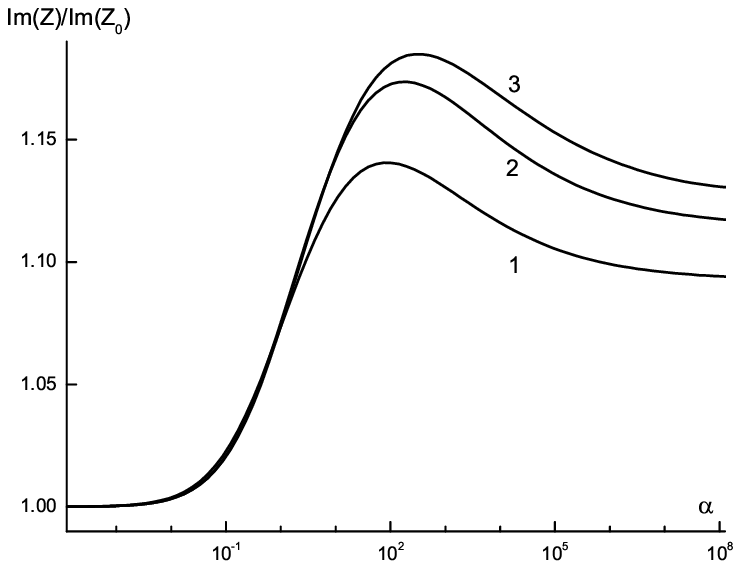}
\caption{Dependence of $\Im(Z)/\Im(Z_0)$ on parameter $\alpha$
for the case $q=0$. The curve $3$ corresponds to diffuse
scattering boundary conditions, the curves $1,2$ correspond to
first and second approximations.}\label{rateII}
\end{center}
\end{figure}


\end{document}